\documentclass{article}

\usepackage{algorithm}
\usepackage{algpseudocode}
\usepackage{balance}

\usepackage{silence}
\usepackage{spconf,amsmath,graphicx,hyperref}

\usepackage[final]{microtype}
\usepackage{url}            
\usepackage{float} 
\usepackage{enumitem}
\usepackage[font=footnotesize,skip=2pt]{caption}

\title{SAFE-IMM: Robust and Lightweight RADAR-Based Object Tracking\\ on Mobile Platforms}
\name{Dnyandeep Mandaokar \qquad Bernhard Rinner}
\address{University of Klagenfurt, Institute of Networked and Embedded Systems\\
Klagenfurt, Austria\\ 
dnyandeep.mandaokar@aau.at, bernhard.rinner@aau.at}

\begin{document}
%
\maketitle
\begin{abstract}
Tracking maneuvering targets requires estimators that are both responsive and robust. Interacting Multiple Model (IMM) filters are a standard tracking approach, but fusing models via Gaussian mixtures can lag during maneuvers. Recent winner-takes-all (WTA) approaches react quickly but may produce discontinuities. We propose SAFE-IMM, a lightweight IMM variant for tracking on mobile and resource-limited platforms with a safe covariance-aware gate that permits WTA only when the implied jump from the mixture to the winner is provably bounded. 
In simulations and on nuScenes front-RADAR data, SAFE-IMM achieves high accuracy at real-time rates, reducing ID switches while maintaining competitive performance. The method is simple to integrate, numerically stable, and clutter-robust, offering a practical balance between responsiveness and smoothness.
\end{abstract}
\section{Introduction}
\label{sec:intro}
Abrupt maneuvers degrade tracking performance with single-model Kalman Filters (KF) and nonlinear filters. The Interacting Multiple Model (IMM) approach addresses this by running multiple motion models and fusing them through a Markovian Transition Probability Matrix (TPM)~\cite{blom1988interacting,genovese2001imm}. Therefore, IMM is widely applied in RADAR, UAV, and multi-sensor systems because it balances accuracy and robustness.

However, IMM fusion is computationally expensive. The Gaussian mixture outputs remain continuous, but at a rather low rate due to the computational complexity, with limited applicability on resource-limited platforms. Lightweight $\alpha$–$\beta$ and $\alpha$–$\beta$–$\gamma$ filters support fast IMM, reducing runtime by more than 60\% compared to KF-IMM while maintaining similar accuracy~\cite{dahmani2010fastIMM}. fastIMM-EV extends this idea by adding Viterbi selection for efficient maneuver recognition~\cite{di2025fastIMM-EV}. Nonlinear variants such as EKF, strong-tracking square-root CKF~\cite{han2019STSRCKF}, or random-weighted CKF~\cite{ma2020ASTRWCKF} improve convergence under abrupt motion. More recently, machine learning-enhanced IMMs with online parameter adaptation in UAV tracking avoid hard tuning~\cite{zeng2022SVRIMM}. These methods improve accuracy but further increase computational cost compared with lightweight filters or linear sub-filter models.

Beyond efficiency, IMM is also sensitive to corrupted measurements. This has motivated robust likelihood designs. Huber-based derivative-free KFs improve resilience to contaminated Gaussian noise~\cite{chang2013robust}, while Student-$t$ formulations provide heavy-tailed robustness in Bayesian filtering and smoothing~\cite{roth2017robust}. 
Another key challenge is TPM adaptation. Lee and Park~\cite{lee2023improvedIMM} proposed dual correction functions to polarize or activate transitions depending on the maneuvering state.
Whereas Gai and Wang~\cite{gai2024adaptiveTPM} combined probability-difference corrections with a time-window discriminator to accelerate switching. Additionally, the structural variants include variable-structure IMM with smoothing (AVSIMMFS), which dynamically prunes model sets while maintaining accuracy~\cite{hu2023AVSIMMFS}.
All these extensions either fuse models (IMM) or enforce hard selection (WTA), risking discontinuities. Yet, none of the above approaches explicitly control the output continuity when switching from mixture fusion to single-model selection. To our knowledge, no prior IMM variant enforces an explicit tolerance on WTA discontinuity. 

This work introduces a safe covariance-aware gate that bounds drift whenever WTA is selected, thereby preserving continuity without modifying IMM internals. This addresses a gap left by existing methods, improving overall performance. Building on this, the proposed SAFE-IMM combines Student-\(t\) likelihoods and lightweight TPM adaptation to deliver a tracker that is faster and more robust than conventional IMM, while maintaining accuracy. In addition, SAFE-IMM is integrated with Global Nearest Neighbor (GNN) association, providing a simple yet efficient end-to-end Multi-Object Tracking (MOT) framework~\cite{smith2019systematic}.

\section{Methodology}
\label{sec:methodology}

\subsection{Conventional IMM}

The conventional IMM with $M$ models $\{\mathcal{M}_i\}_{i=1}^M$, each producing a Gaussian posterior $\mathcal{N}(\mu_i,P_i)$ after correction, returns the single-Gaussian mixture~\cite{genovese2001imm}
\begin{equation}
\label{eq:mm}
\begin{aligned}
\mu_{\mathrm{mix}}&=\sum_{i=1}^M c_i\,\mu_i ,\\
P_{\mathrm{mix}}&=\sum_{i=1}^M c_i\!\Big(P_i+(\mu_i-\mu_{\mathrm{mix}})(\mu_i-\mu_{\mathrm{mix}})^\top\Big) ,
\end{aligned}
\end{equation}
where $c_i=\Pr(\mathcal{M}_i\mid\mathcal{Z}_{1:k})$ denotes the posterior model probabilities at time $k$.
Model priors are $c_{k|k-1}=\Pi^\top c_{k-1|k-1}$; mixing into model $j$ uses $\bar c_j=\sum_{i}\Pi_{ij}c_i$ and weights $\alpha_{i\to j}=\frac{\Pi_{ij}c_i}{\bar c_j}$, yielding mixed $(\mu_{0,j},P_{0,j})$ via the standard IMM formulas (state-space mapped by $T_{i\to j}$).
Each model likelihood under Gaussian innovation is \cite{genovese2001imm}
\begin{align}
\label{eq:like}
\mathcal{L}_j &\propto |S_j|^{-1/2}
   \exp\!\left(-\tfrac12\,v_j^\top S_j^{-1}v_j\right) , & \\
S_j &= H_j P_j^- H_j^\top + R_j ,  &
\end{align}
\noindent where $S_j$, $H_j$, and $R_j$ denote the innovation covariance, measurement matrix, and measurement noise covariance of model $j$, respectively. 

\subsection{SAFE-IMM}
We integrate a post-fusion, covariance-aware gate that provably bounds the drift when the output follows a WTA model. In addition, the method hardens the pipeline with robust Student-\(t\) likelihoods and simple TPM adaptations based on generalized likelihood ratio (GLR), entropy, and winner streak.

Let $w = c_{k|k}$ denote the posterior model-probability vector, with 
$w^\star = \arg\max_i w_i$ the index of the winning model, 
$w_w = \max_i w_i$ the winning probability, 
and the tail mass $t = 1 - w_w$.
Map each rival model $i\neq w^\star$ into the winner’s state space via $T_{i\to w^\star}$ defined as
\begin{flalign}
\mu_{i\to w} &= T_{i\to w^\star}\mu_i , \\
P_{i\to w} &= T_{i\to w^\star}P_iT_{i\to w^\star}^\top , \\
\Delta_i &= \mu_{i\to w}-\mu_w ,
\end{flalign}

\noindent where $(\mu_w,P_w)$ is the winner’s mean and covariance, with $w$ standing for $w^\star$. If $t>0$, the normalized rival weights are defined as $\tilde w_i = w_i/t$ (so $\sum_{i\neq w}\tilde w_i=1$). The reference covariance matrix $\bar P$ and weighted Mahalanobis spread of rival means $\overline{d^2}$ are defined as
\begin{flalign}
\label{eq:Pbar}
\bar P &= \tfrac12\!\Big(P_w+\sum_{i\neq w}\tilde w_i\,P_{i\to w}\Big) , \\
\overline{d^2} &= \sum_{i\neq w}\tilde w_i\,\Delta_i^\top \bar P^{-1}\Delta_i .
\end{flalign}

We assume $\bar P\succ0$ so that $\bar P^{-1}$ exists. Applying the Cauchy–Schwarz inequality in the $\bar P^{-1}$ inner product yields the following rank-one PSD (Loewner) inequality
\[
vv^\top \preceq (v^\top \bar P^{-1} v)\,\bar P\quad(\text{for all }v),
\]
which yields the drift bound (Euclidean norm)
\begin{equation}
\label{eq:driftbound}
\big\|\mu_{\mathrm{mix}}-\mu_w\big\|
\ \le\ B\ =\ t\,\sqrt{\operatorname{tr}(\bar P)\,\overline{d^2}} .
\end{equation}

If $B\le \varepsilon_{\mathrm{WTA}}$, we permit WTA and output $(\mu_w,P_w)$; else we output the conventional mixture~\eqref{eq:mm}.
Thus, WTA is permitted whenever (WTA gate)
\begin{equation}
\label{eq:guarantee}
\big\|\mu_{\mathrm{mix}}-\mu_w\big\|\le \varepsilon_{\mathrm{WTA}} .
\end{equation}

\begin{figure}
  \centering
  \includegraphics[width=\linewidth]{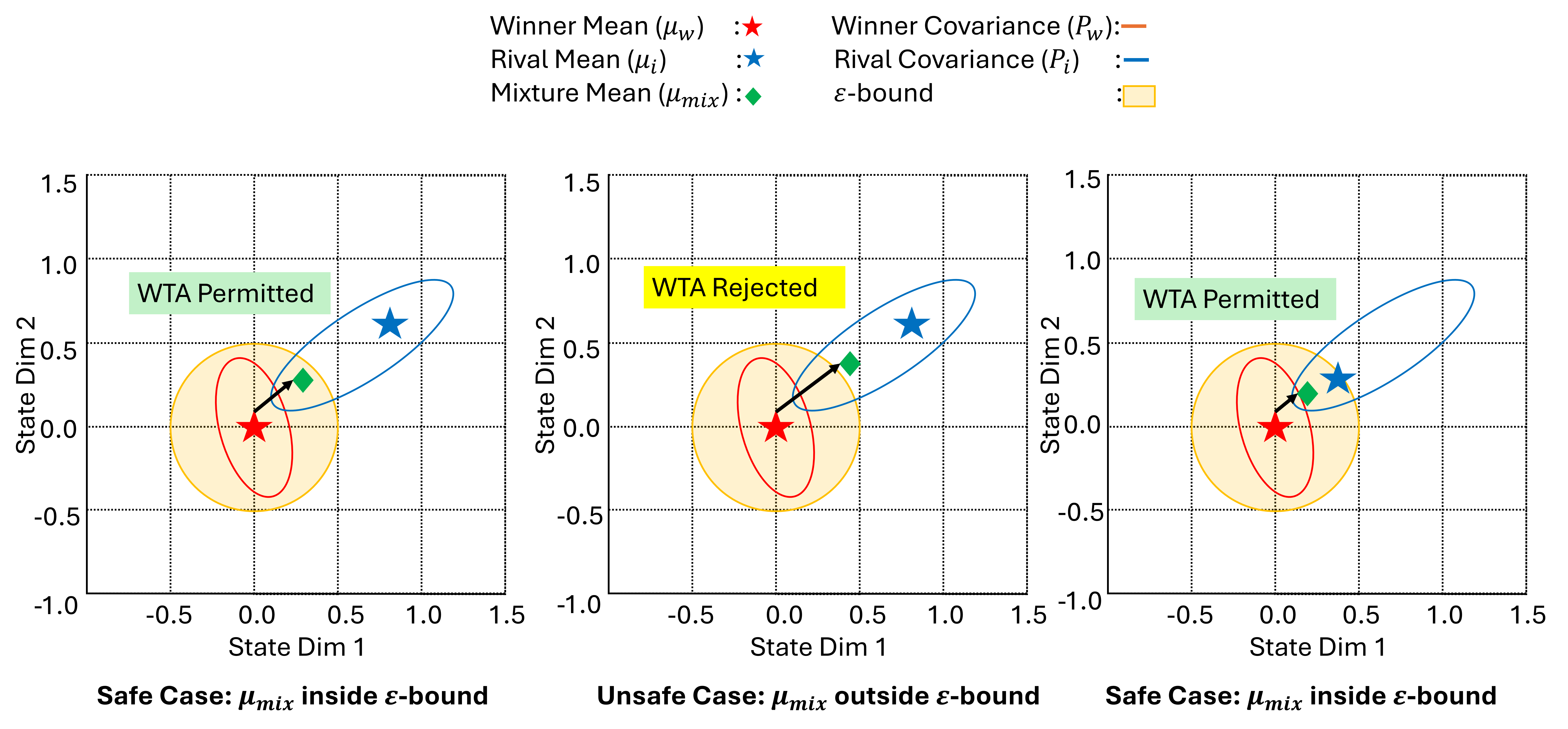}
  \caption{Illustration of the SAFE WTA rule with covariance geometry. 
  The panels demonstrate when WTA is permitted (left), rejected (center), or permitted with a close rival (right).}
  \label{fig:esafebound}
\end{figure}

Fig.~\ref{fig:esafebound} left and middle show the WTA permitted and rejected cases, and Fig.~\ref{fig:esafebound} right presents a close rival case. This demonstrates how the WTA gate not only accounts for the mean distances between 
the mixture and the winner, but also for the underlying covariance geometry 
(Mahalanobis distance) that shapes the safe boundary. 
In particular, even when a rival mean appears close in Euclidean space, an elongated covariance can imply a large Mahalanobis separation. Thus, the bound guarantees that WTA is permitted only when both the probability mass and the geometry of uncertainty will remain within the tolerance 
$\varepsilon_{\mathrm{WTA}}$.  In such cases, its cost approaches that of a single-model KF, while retaining IMM-level robustness when WTA is not triggered. Algorithm~1 outlines one iteration of our SAFE-IMM:

\begin{algorithm}[H]
\caption{One IMM iteration with $\varepsilon$-gate SAFE WTA}
\begin{algorithmic}[1]
\State $c_{k|k-1}=\Pi^\top c_{k-1|k-1}$ \hfill (priors)
\For{$j=1,\dots,M$}
  \State Mix into $j$: $(\mu_{0,j},P_{0,j})$; predict \& update;
  \State Compute $v_j,S_j$
  \State Likelihood $\mathcal{L}_j$ (Gaussian or robust) \eqref{eq:like}
\EndFor
\State Posterior $w=c_{k|k}$; identify winner $w^\star$, $t=1-\max_i w_i$
\State Compute $(\mu_{\mathrm{mix}},P_{\mathrm{mix}})$ via \eqref{eq:mm}
\State Build $\bar P$, $\overline{d^2}$; compute $B$ \eqref{eq:driftbound}
\If{$B\le \varepsilon_{\mathrm{WTA}}$}
  \State \textbf{Output} $(\mu_w,P_w)$ \Comment{WTA fires; mean drift $\le\varepsilon_{\mathrm{WTA}}$}
\Else
  \State \textbf{Output} $(\mu_{\mathrm{mix}},P_{\mathrm{mix}})$ \Comment{No WTA}
\EndIf
\end{algorithmic}
\end{algorithm}

\subsection{Hardening Strategies}
In addition to the theoretical SAFE WTA gate, the SAFE-IMM integrates the following modifications that improve robustness and stability. \textbf{Data Association:} There are various methods for tracking, such as Joint Probabilistic Data Association (JPDA) or PDA, and the Multiple Hypothesis Theorem (MHT). The JPDA and MHT are effective in cluttered environments but are much slower and computationally heavier ~\cite{ defeo1997immjpda}. Thus, to support the aim of our method for lightweight tracking, we use a GNN tracker for association with SAFE-IMM to perform multi-object tracking. \textbf{Robust Likelihoods:} We replace the Gaussian likelihood
$\mathcal{L}_j$ in~\eqref{eq:like} with robust Student-$t$.
These heavy-tailed distributions down-weight large residuals, tempering
the effect of clutter and outliers~\cite{chang2013robust,roth2017robust}.
\textbf{Adaptive Transition Matrix:} Instead of a fixed TPM, we apply online corrections: a probability cap, GLR/entropy-based blending, winner-streak self-transition bias, and small boosts for CA/CV maneuver models (cf.~\cite{gai2024adaptiveTPM,hu2023AVSIMMFS,lee2023improvedIMM}). These adjustments accelerate switching to the appropriate model while maintaining robustness to noise.


\section{Experiments}
\label{sec:experiment}
The SAFE-IMM is assessed in simulations under maneuvers and noise. Additionally, it is validated on the nuScenes validation split~\cite{caesar2020nuscenes} using the front-RADAR (70$^\circ$ FoV, $\leq$70\,m) with dynamic-only ground truth from the aligned front camera. Comparisons are made against conventional IMM baselines, linear KF, and nonlinear EKF with Constant Velocity (CV), Constant Acceleration (CA), Constant Turnrate (CT) models, and published monocular front-view trackers~\cite{zhou2020centertrack,wu2021trades,huang2021delving}. Against these baselines, SAFE-IMM employs only two KFs (CV and CA) to remain lightweight without sacrificing robustness. Because CV and CA differ in state size, rivals are mapped to the winner's space.
The baseline IMMs and SAFE-IMM are combined with the GNN tracker. All runs were executed on a CPU-only laptop (Intel Core Ultra 7 155H @1.40\,GHz, 32\,GB RAM, 64-bit OS), ensuring results reflect real-time performance without GPU support. The model is designed and validated in MATLAB. All parameters discussed further were fixed across all experiments.

\subsection{Simulation Setup}
\label{ssec:simsetup}
The proposed SAFE IMM is tested and evaluated in MOT scenarios. The simulation runs $\Delta t=0.1\,\text{s}$ with total durations of $T=30\,\text{s}$. We use two models CV ($n_x=6$) and CA ($n_x=9$) with process noise intensities $q_{\text{CV}}=0.5$ and $q_{\text{CA}}=0.2$.
The base TPM is  
\[
\Pi_{\text{base}} =
\begin{bmatrix}
0.992 & 0.008 \\
0.015 & 0.985
\end{bmatrix},
\]
which is adaptive TPM using GLR- and entropy-based blending with $\alpha_{\max}=0.7$, $g_{\text{GLR}}=0.10$, $g_{\text{Ent}}=0.50$, winner bias $0.10$, CA boost $0.15$, CV boost $0.05$. Robust likelihoods are enabled via Student-$t$ ($\nu=5$) with stronger tail under jamming ($\nu_{\text{jam}}=2$). For gating, we set $\varepsilon_{\text{WTA}}=0.5$, margin $\delta=0.05$, and hysteresis streak length $2$. MOT tracks three targets with: (T1) with nearly constant velocity, (T2) with moderate maneuvers to assess switching responsiveness, and (T3) with sharp accelerations to stress the robustness of SAFE-IMM under aggressive dynamics. Tracking follows standard GNN practice: assignment threshold 30, deletion after 3 missed detections, and confirmation thresholds [2, 5].

\noindent\textbf{Evaluation Metrics:} 
The Root-Mean-Square Error (RMSE) is computed for state estimation accuracy.
For MOT performance, the Optimal Sub-Pattern Assignment (OSPA) distance quantifies both localization error and cardinality mismatch. The details of the used RMSE and OSPA metric are found in ~\cite{ma2020ASTRWCKF,ristic2011metric}.

\subsection{Simulation Result}
\label{ssec:simresult}

\begin{figure}
  \centering
  \includegraphics[width=\linewidth, height= 3.5cm]{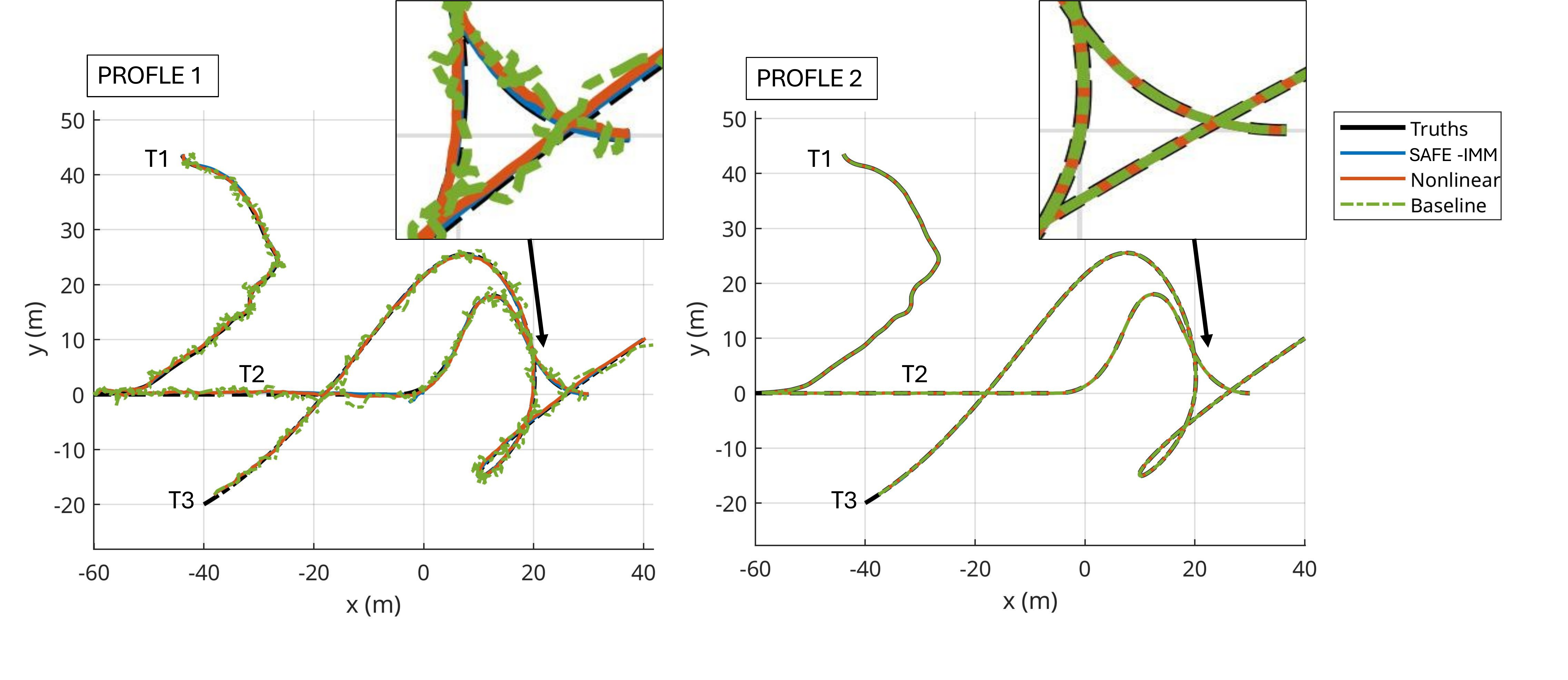}
  \caption{Ground truth vs.~simulated tracks of targets T1-T3 under Profiles 1 (high position noise) and 2 (high velocity noise).}
  \label{fig:alltracks}
\end{figure}

The SAFE-IMM is evaluated with two noise profiles for Profile 1 (2.0$\, $m, 0.01$\, $m/s) and Profile 2 (0.01$\, $m, 2.0$\, $m/s). For Profile 1, SAFE-IMM yields the lowest mean RMSE $[0.135,0.179]\, $m (pos) and $[0.120,0.014]\, $m/s (vel), mean OSPA \(1.070\); and follows the same trend for localization \(0.292\) and cardinality \(0.667\). It outperforms nonlinear and baseline IMMs in position RMSE and OSPA  with \(100\%\) $\varepsilon$-gate compliance. Table~\ref{tab:mot_rmse_ospa} summarizes per-target scores and Fig.~\ref{fig:alltracks} shows MOT tracks for both profiles.

\begin{table}
\centering
\footnotesize
\setlength{\tabcolsep}{4pt}
\renewcommand{\arraystretch}{1.05}
\caption{Profile 1 Per-target $x,y$ RMSE and OSPA.}
\label{tab:mot_rmse_ospa}
\begin{tabular}{c|c|c|c}
\hline
\textbf{Target} & \textbf{Tracker} & \textbf{RMSE$_{x,y}$ [m] $\downarrow$} & \textbf{OSPA (mean\,|\,loc\,|\,card) $\downarrow$} \\
\hline
T1 & SAFE-IMM  & [0.173, 0.242] & 0.994 \,|\, 0.327 \,|\, 0.667 \\
   & Nonlinear & [0.230, 0.237] & 1.034 \,|\, 0.367 \,|\, 0.667 \\
   & Baseline  & [0.819, 0.714] & 1.963 \,|\, 1.297 \,|\, 0.667 \\
\hline
T2 & SAFE-IMM  & [0.125, 0.358] & 1.334 \,|\, 0.334 \,|\, 1.000 \\
   & Nonlinear & [0.218, 0.303] & 1.064 \,|\, 0.398 \,|\, 0.667 \\
   & Baseline  & [0.653, 0.820] & 1.988 \,|\, 1.321 \,|\, 0.667 \\
\hline
T3 & SAFE-IMM  & [0.108, 0.118] & 0.884 \,|\, 0.217 \,|\, 0.667 \\
   & Nonlinear & [0.160, 0.264] & 1.003 \,|\, 0.337 \,|\, 0.667 \\
   & Baseline  & [0.707, 0.740] & 1.926 \,|\, 1.259 \,|\, 0.667 \\
\hline
\end{tabular}
\end{table}

\begin{table}
\centering
\footnotesize
\setlength{\tabcolsep}{4pt}
\renewcommand{\arraystretch}{1.05}
\caption{Profile 2 Per-target $x,y$ RMSE and OSPA.}
\label{tab:mot_rmse_ospa_new}
\begin{tabular}{c|c|c|c}
\hline
\textbf{Target} & \textbf{Tracker} & \textbf{RMSE$_{x,y}$ [m] $\downarrow$} & \textbf{OSPA (mean\,|\,loc\,|\,card) $\downarrow$} \\
\hline
T1 & SAFE-IMM  & [0.010, 0.009] & 0.681 \,|\, 0.014 \,|\, 0.667 \\
   & Nonlinear & [0.010, 0.009] & 0.682 \,|\, 0.015 \,|\, 0.667 \\
   & Baseline  & [0.010, 0.009] & 0.682 \,|\, 0.015 \,|\, 0.667 \\
\hline
T2 & SAFE-IMM  & [0.025, 0.010] & 0.685 \,|\, 0.020 \,|\, 0.667 \\
   & Nonlinear & [0.009, 0.010] & 0.682 \,|\, 0.016 \,|\, 0.667 \\
   & Baseline  & [0.009, 0.010] & 0.682 \,|\, 0.016 \,|\, 0.667 \\
\hline
T3 & SAFE-IMM  & [0.017, 0.016] & 0.686 \,|\, 0.022 \,|\, 0.667 \\
   & Nonlinear & [0.010, 0.011] & 0.683 \,|\, 0.016 \,|\, 0.667 \\
   & Baseline  & [0.010, 0.011] & 0.683 \,|\, 0.016 \,|\, 0.667 \\
\hline
\end{tabular}
\end{table}

\begin{figure}
  \centering
  \includegraphics[width=\linewidth]{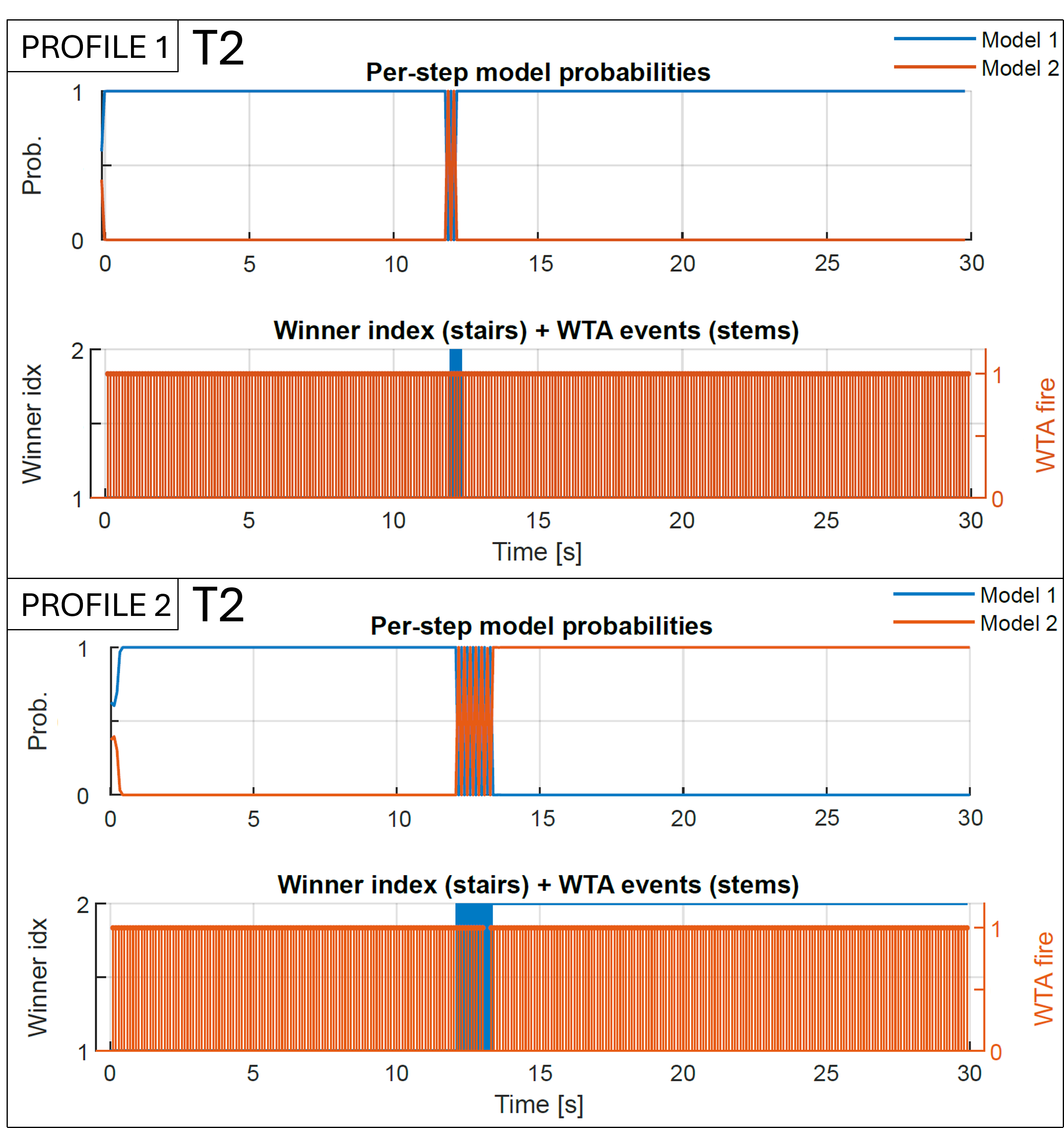}
  \caption{Model probabilities (CV:1, CA:2) and WTA decisions for T2. The orange spike shows WTA fire, and the blue spike shows the winner change.}
  \label{fig:modelprob}
\end{figure}

Under Profile 2, mean RMSE close to \([0.017,0.012]\)\,m (pos) and \([0.180,0.117]\)\,m/s (vel), with mean OSPA of 0.683 and 100\% $\varepsilon$-gate compliance. 
The RMSE for T2 and T3 is higher than that of baseline models. When velocity noise dominates, SAFE-IMM’s conservative WTA triggers more switches and sacrifices a bit of RMSE (see Fig.~\ref{fig:modelprob}). The OSPA metrics for baseline, nonlinear, and SAFE-IMM are close. SAFE-IMM has better localization and cardinality for T1 and slightly lower for T2 and T3 against the other two trackers. Table~\ref{tab:mot_rmse_ospa_new} reports per-target values. In Fig.~\ref{fig:modelprob}, T2 shows the per-step model probabilities, winner index, and WTA events in SAFE-IMM. For Profile 1, Model 1 dominates with only brief switching near \(t{\approx}11\,\)s; the blue stem markers indicate WTA fires, which consistently selected Model 1 in \(299/300\) frames (min margin 1.99). In Profile 2, more frequent switching occurs around \(t{\approx}10\)-15\,s, where Model 2 intermittently wins, resulting in higher RMSE; the WTA mechanism still fired in \(299/300\) frames, ultimately committing to Model 2 (min margin 1.62), ensuring 100\% compliance.


\noindent\textbf{Ablation Study}: The IMM without the SAFE gate yields mean RMSEs of \([0.133,4.929]\)\,m (pos), with mean OSPA $2.895$, notably degraded on T2. With the SAFE gate, performance is markedly improved (see Table~\ref{tab:mot_rmse_ospa}). We also evaluated the impact of the SAFE gate under two likelihood models with higher noise of (0.30$\,$m, 8.00$\,$m/s). Gaussian showed larger bound excursions of max $22.5$ against Student-$t$, which was more robust to outliers, with max $18.3$. The SAFE gate for Profile 1 shows that the fixed TPM yielded slightly lower mean OSPA (0.975 vs.\ 1.080) and marginally lower velocity RMSEs, while the adaptive TPM improved T2 position along $x/y$ (0.148/0.143 vs.\ 0.270/0.199)\,m. The fixed TPM suffices, reserving adaptive TPM for maneuver-rich scenarios.

\subsection{Evaluation with NuScenes Dataset}
\label{ssec:nuscenes}

\begin{table}
\caption{NuScenes Front-RADAR validation.}
\centering
\label{tab:main_results}
\vspace{0.50em}
\scriptsize
\setlength{\tabcolsep}{4pt}
\renewcommand{\arraystretch}{1.2}
\resizebox{\columnwidth}{!}{%
\begin{tabular}{|l|c|c|c|c|c|c|c|}
\hline
\textbf{Model} &
\begin{tabular}[c]{@{}c@{}}AMOTA\\(\%)$\uparrow$\end{tabular} &
\begin{tabular}[c]{@{}c@{}}AMOTP\\(m)$\downarrow$\end{tabular} &
\begin{tabular}[c]{@{}c@{}}MOTA\\(\%)$\uparrow$\end{tabular} &
\begin{tabular}[c]{@{}c@{}}MOTP\\(m)$\downarrow$\end{tabular} &
\begin{tabular}[c]{@{}c@{}}MOTAR\\$\uparrow$\end{tabular} &
\begin{tabular}[c]{@{}c@{}}Recall\\$\uparrow$\end{tabular} &
\begin{tabular}[c]{@{}c@{}}IDS\\$\downarrow$\end{tabular} \\
\hline
SAFE-IMM   & \textbf{7.24} & 0.178 & \textbf{18.5} & 1.261 & \textbf{0.480} & 0.386 & \textbf{37} \\
Nonlinear  & 6.01 & \textbf{0.146} & 17.1 & \textbf{1.150} & 0.427 & \textbf{0.400} & 91 \\
Baseline   & 6.28 & 0.169 & 16.1 & 1.208 & 0.409 & 0.395 & 105 \\
\hline
\end{tabular}}
\end{table}

For the evaluation, only dynamic ground-truth objects aligned with the front-facing camera were considered, consistent with the nuScenes annotation protocol~\cite{caesar2020nuscenes}. Following the MoMA-M3T protocol~\cite{huang2021delving, caesar2020nuscenes}, performance is reported using class-agnostic AMOTA/AMOTP. SAFE-IMM+GNN uses a sliding-window RADAR object detector~\cite{mandaokar2025mmwave} to keep the pipeline efficient. SAFE-IMM surpasses both conventional IMM variants, improving AMOTA/MOTA and sharply reducing IDS switches by ~60\% (see Table~\ref{tab:main_results}), while running in real time. Compared to monocular front-view baselines on the same split (CenterTrack~\cite{zhou2020centertrack}: 4.6\% AMOTA; TraDeS~\cite{wu2021trades}: 5.9\%; PermaTrack~\cite{huang2021delving}: 6.6\%), SAFE-IMM achieves higher AMOTA. Although the sensing modalities differ (camera vs RADAR), the numbers are provided only for context and show a modest increase in AMOTA for the RADAR-only modality, with SAFE-IMM at 1.0–1.5\% higher than for camera-only tracking. SAFE-IMM yields slightly lower recall because its covariance-aware gating and conservative WTA suppress marginal tracks, trading coverage for greater consistency, robustness, and ID stability. All models share the same GT set ($\sim$13.8k objects). SAFE-IMM achieves 17.9\,FPS end-to-end and 229.6\,FPS IMM-only (Baseline: 9.3\,FPS/198.1\,FPS; Nonlinear: 8.8\,FPS/187.9\,FPS).

\section{Conclusion}
\label{sec:Conclusion}
SAFE-IMM combines a covariance-aware gate, robust likelihoods, and adaptive TPM to guarantee bounded drift under WTA. Simulations and nuScenes experiments demonstrate accuracy, robustness, and real-time performance compared with conventional and nonlinear IMM baselines. The SAFE-IMM is computationally inexpensive with a WTA single model output for interaction and prediction. SAFE-IMM is a practical, principled tracker for maneuvering targets under resource constraints. The multi-sensor and category-aware extensions are left to future work.

\vfill\pagebreak


\bibliographystyle{IEEEbib}
\bibliography{main}

\end{document}